\documentstyle[12pt]{article}
\newdimen\lineskp
\newdimen\arrayskp
\newcommand{\mabel}[1]{\label{#1}} 
\newcommand{\mibitem}[1]{\bibitem{#1}}

\renewcommand{\slash}{~/\!\!\!\!}
\newcommand{\be}{\begin{equation}}
\newcommand{\ee}{\end{equation}}
\newcommand{\bea}{\begin{eqnarray}}
\newcommand{\eea}{\end{eqnarray}}
\newcommand{\beastar}{\begin{eqnarray*}}
\newcommand{\eeastar}{\end{eqnarray*}}

\newcommand{\trac}[2]{{\textstyle\frac{#1}{#2}}}
\newcommand{\abs}[1]{\left| #1 \right|}

\newcommand{\e}{{\rm e}}
\newcommand{\Det}{{\rm Det}}
\newcommand{\Tr}{{\rm Tr}}
\newcommand{\Ad}{{\rm Ad}}
\newcommand{\del}{\partial}
\newcommand{\N}{{\cal N}}
\renewcommand{\P}{{\cal P}} 
\renewcommand{\O}{{\cal O}} 
\newcommand{\X}{{\cal X}}
\newcommand{\Z}{{\cal Z}}

\newcommand{\unit}{1\!\!\hskip 1pt\mbox{l}}

\def\sCC{{\kern 0.27em\vrule height1.45ex width0.03em depth0em
          \kern-0.30em\rm C}}
\def\CC{{\mathchoice
  {\sCC}
  {\sCC}
  {\kern 0.225em \vrule height1.05ex width0.025em depth0em \kern-0.25em \rm C}
  {\kern 0.180em \vrule height0.78ex width0.02em depth0em \kern-0.2em \rm C}}}
\def\sRR{{\rm I\kern-0.16em{}R}}
\def\RR{{\mathchoice
  {\sRR}
  {\sRR}
  {\rm I\kern-0.12em{}R}
  {\rm I\kern-0.10em{}R} }}
\def\sZZ{{\rm Z\kern-0.32em{}Z}}
\def\ZZ{{\mathchoice
  {\sZZ}
  {\sZZ}
  {\rm Z\kern-0.30em{}Z}
  {\rm Z\kern-0.25em{}Z} }}

\newcommand{\NPB}{{~Nucl. Phys.} B~}
\newcommand{\PRL}{{~Phys. Rev. Lett. }}
\newcommand{\PRD}{{~Phys. Rev.} D~}

\newcommand{\PLB}{{~Phys. Lett.} B~}
\newcommand{\PLA}{{~Phys. Lett.} A~}
\newcommand{\PR}{{~Phys. Rep. }}

\newcommand{\CMP}{{~Commun. Math. Phys. }}

\newcommand{\MPLA}{{~Mod. Phys. Lett.} A~}

\newcommand{\IJMPB}{{~Int. J. Mod. Phys.} B~}
\newcommand{\AP}{{~Ann. Phys. (NY)~}}

\newcommand{\TMP}{{~Theor. Math. Phys. }} 

\begin{document}
\baselineskip= \lineskp plus 1pt minus 1pt


\pagestyle{empty}

\rightline{UBC-GS-96-3}
\rightline{hep-th/9609228}
\rightline{   }
\rightline{\today}

\vskip 0.5truein
\begin{center}

{\Large\bf G/G models as the strong coupling limit of topologically
massive gauge theory}\\
\vskip 0.5truein
{\bf G. Grignani$^a$, G. Semenoff$^b$, P. Sodano$^a$ and O.
Tirkkonen$^b$}\\
\medskip
{\it $(a)~$Dipartimento di Fisica e Sezione I.N.F.N.\\
Universit\'a di Perugia, Via A. Pascoli\\
I-06123 Perugia, Italia\\~~
\\$(b)~$Department of Physics and Astronomy,\\
University of British Columbia\\6224 Agricultural Road\\
Vancouver, British Columbia, Canada V6T 1Z1}\\

\bigskip
\vfill

{\bf Abstract}

\end{center}


\noindent
We show that the problem of computing the vacuum expectation values
of gauge invariant operators in the strong coupling limit of
topologically massive gauge theory is equivalent to the problem of
computing similar operators in the $G_k/G$ model where $k$ is the
integer coefficient of the Chern-Simons term.  The $G_k/G$ model is
a topological field theory and many correlators can be computed
analytically.  We also show that the effective action for the
Polyakov loop operator and static modes of the gauge fields of the
strongly coupled theory at finite temperature is a perturbed,
non-topological $G_k/G$ model.  In this model, we compute the one loop 
effective potential for the Polyakov loop operators and explicitly 
construct the low-lying excited states. In the strong coupling limit
the theory is in a deconfined phase.


\newpage \pagestyle{plain} \setcounter{page}{1}   


\section{Introduction}

Chern-Simons theory \cite{tpm,djt} is a three dimensional topological
quantum field theory whose observables and correlation functions are
intimately related the topology of three dimensional manifolds
\cite{witt3}.  It has found physical applications to quasi
2+1-dimensional systems where some of the topological effects which
are associated with it are considered important.  A well known example
is the use of both abelian and non-abelian Chern-Simons theory to
describe the the quantum Hall states.  There, the exotic statistics of
quasiparticles \cite{frstat}
which arise from their coupling to Chern-Simons theory
is an essential feature.  Another example are speculations about the
mechanism for high temperature superconductivity \cite{hiT}. A very
interesting recent application is the relation of Chern-Simons theory
to quark-gluon plasmas in four dimensions \cite{en}.

The Chern-Simons action is the volume integral of a three-form
\begin{equation}
S_{CS}=\frac{k}{4\pi}\int d^3x~{\rm Tr}\left( AdA-\frac{2}{3}iA^3\right)
 \mabel{CS}
\end{equation}
where $A$ is the connection one-form and $k$ is an integer.  This
action involves no dimensional parameters.  Furthermore, being the
volume integral of a three-form, it does not contain the metric of the
three dimensional space and it is therefore invariant under general
coordinate transformations.  This is a characteristic feature of a
topological field theory.  Assuming that the integration measure can
be appropriately defined, the partition function
\begin{equation}
Z=\int [DA]~e^{iS_{CS}[A]}
\end{equation}
is independent of local geometry and depends only on topological data.

In physical applications, coupling of the gauge field
to other degrees of freedom of a
physical system usually breaks the general coordinate invariance of
Chern-Simons theory so that it is no longer a topological field
theory. Generally, the leading correction
to the effective action for the gauge field
in powers of gauge fields and their derivatives
is a Yang-Mills term,\footnote{For certain kinds of matter fields, there could
also be renormalization of the Chern-Simons term \cite{rns}.}
\begin{equation}
S_{YM}=-\int d^3x ~\frac{1}{2e^2}
{\rm Tr}\left( F_{\mu\nu}F^{\mu\nu}\right)
\mabel{YM}
\end{equation}
where $F$ is the field strength and $e^2$ is a parameter
with the dimension of mass.  Such a term would be generated by
radiative corrections from relativistic matter fields with a mass gap.  The
coupling constant $e^2$ is typically of the same magnitude as the mass gap.
In this Paper, we shall consider this theory in the limit where the dimensional
constant $e^2$ is large compared to other momentum scales of interest.

If the matter fields which the gauge field coupled  were not relativistically
invariant, but had a non-relativistic, spatially isotropic spectrum, the induced
Yang-Mills action would have two independent parameters,
\begin{equation}
S_{YM}^{{\rm ~nr}}=-\int d^3x ~{\rm Tr}\left(
\frac{1}{2e^2}F_{0i}F^{0i}+ \frac{1}{2P_M}F_{ij}F^{ij}\right)
\mabel{YMnr}
\end{equation}
with $e^2$ being related to the electric permeability and
$P_M$ the magnetic
permittivity.  The strong coupling limit which we shall consider in
this paper applies to the situation
where
\begin{equation}
1/e^2>>1/P_M
\mabel{limit}
\end{equation}
or where $P_M$ is much larger than any other dimensional parameters of
interest, so that the
electric term dominates and the magnetic term can be ignored entirely.
This is typical in the case where the average velocities and effective
charges of the non-relativistic matter are small so that their motions are
unaffected by magnetic fields.  If (\ref{limit}) holds, the analysis which we
shall describe in this
paper is valid without the constraint that $e^2$ also be large.
In the following, for the
purposes of discussion, we shall assume that the Yang-Mills action has the
relativistic form (\ref{YM}) and take the strong coupling, large $e^2$ limit.

The presence of the Yang-Mills term in the action breaks the general
coordinate invariance of the Chern-Simons term explicitly.  It also
renders the gauge interaction super-renormalizable.  It could be
introduced for that purpose, as a higher derivative ultraviolet
regulator.  In that case $e^2$ would be of order the ultraviolet
cutoff.

Topologically massive Yang-Mills theory, which is the model with both
the Chern Simons and Yang Mills terms in the action, \cite{tpm,djt},
\begin{equation}
S=S_{YM}+S_{CS}
\end{equation}
describes a self-interacting massive vector field in 2+1-dimensions.
The coupling is weak when the integer $k$ is large, which also means
that the vector field mass $\mu=e^2k/4\pi$ is large.

In this Paper, we shall make some observations concerning the strong
coupling limit, $e^2\rightarrow\infty$,
with $k$ fixed and not
necessarily large, of topologically massive gauge
theory\footnote{Since $e^2$ is a dimensionful parameter, what is meant
by this limit is that $e^2$ should be large compared to all momentum
scales of interest.  Later at finite temperature, the appropriate
limit will be that $e^2T\rightarrow\infty$ in such a way that $e^2/T$
remains finite.}.  That limit should recover Chern-Simons theory with
some integer parameter $k'$ (not necessarily the original classical
$k$).  We shall show that, in the Hamiltonian formalism, it is also
equivalent to another type of topological field theory, the G/G model,
which is also known (by other means \cite{spiegel,witt1,bt1,geras}) to
be related to Chern-Simons theory.

Typically, the strong coupling limit of a gauge theory is not a
renormalizable field theory.  It is interesting that topologically
massive gauge theory avoids this fate by making use of general
coordinate invariance, i.e. that the strongly coupled theory itself is
a topological field theory.  From a physical point of view, in the
strong coupling limit, the mass gap of the theory is taken to
infinity, leaving only dynamics of the vacuum which involves no
propagating field theoretical degrees of freedom which could produce
ultraviolet divergences.

The resulting quantum mechanical degrees of
freedom are described by the G/G model where, in principle, many
interesting correlators can be computed exactly.  We shall show that
the problem of computing the ground state expectation value of a wide
array of static gauge invariant operators in the strong coupling limit
of 2+1-dimensional topologically massive gauge theory is equivalent to
the problem of computing similar observables in the $G_k/G$ model
where $k$ is the integer quantized coefficient of the Chern-Simons
term in the topologically massive gauge theory.

The $G_k/G$ models \cite{spiegel,witt1,bt1,geras,witt2,bt2,as}
themselves are an interesting class of topological field theories
related to two dimensional Yang-Mills theory.  They are obtained by
taking a conformal field theory, the Wess-Zumino-Novikov-Witten (WZNW) model
\cite{novi,witt4}, whose dynamical variables are loop group elements,
$g(x)\in G$, and introducing gauge fields so that the model is
symmetric under the gauge transformation
\begin{equation}
g(x)\rightarrow u^{-1}(x)g(x)u(x)
~~,~~
A(x)\rightarrow u^{-1}(x)A(x)u(x)+iu^{-1}(x)du(x)
\end{equation}
The gauge fields are integrated over without a kinetic term, so that
they act effectively as Lagrange multipliers. With the gauge field
taking values in an anomaly-free subgroup of $G$, the gauged WZW model
reproduces the coset construction of conformal field theories
\cite{gk}. When the gauge fields take values in whole $G$, the result
is a topological field theory which is also a $c=0$ conformal field
theory.  The conformal field theory has been used to compute the
Verlinde formula which relates the topology of group theoretical
numbers to the dimensions of conformal blocks \cite{bt1,geras}.  Here
we shall use the Verlinde formula to show that the center-symmetry of
topologically massive gauge theory is spontaneously broken in infinite
volume, consistent with the deconfined nature that one would expect for a
topologically massive gauge theory.

A more complex issue is to understand the nature of excited states in
the strong coupling limit.  In a conventional gauge theory,
3+1-dimensional Yang-Mills theory for example, in the strong coupling
limit, excited states contain static, infinitely massive strings of
electric flux.  In a topologically massive gauge theory, they turn out
to be quite different objects.  The spectrum is similar to a tower of
Landau levels, with excited states being occupation of the higher
levels.  Here, we will consider their contribution to
thermodynamic states by constructing an effective field theory for the
finite temperature strong coupling model.  Indeed, to have nontrivial
thermodynamics, the temperature will be taken as large enough to
excite some of the states of the underlying theory.  That is, the
finite temperature theory is no longer Chern-Simons theory but the
true strong coupling limit of topologically massive Yang-Mills theory
where some of the massive degrees of freedom are excited.  We shall
find that the partition function is described by a non-renormalizable
effective field theory, which is a perturbed G/G model.  The degrees
of freedom in this effective model are the static, magnetic modes of
the vector gauge field and the group valued field whose trace is the
Polyakov loop operator.

This model is a topological field theory at the classical level, but
the general coordinate invariance is broken by anomalies when the
model is quantized.  In fact, for arbitrary parameters it is
non-renormalizable.  However, we will find that, when it is tuned so
that it is logarithmically close to a topological field theory, that is 
in the limit where the dimensionless ratio of coupling constant to 
temperature is tuned as
$e^2/T\sim ~\ln\Lambda/\mu$ (with $\Lambda $ the ultraviolet cutoff)
the effective potential for the group valued field can be made finite to one 
loop. In this limit, we then have some indication that the theory is 
one-loop renormalizable.
Then, in principle, correlators of static operators such as 
the Polyakov loop operators and moments of the magnetic field 
can be computed to one-loop order.

We shall also examine the nature of excited states in the strong 
coupling limit.
We shall find that the excited states themselves can be
constructed by using holomorphic factorization properties of the G/G
model.  We will comment on their properties.

In Section 2 we review the quantization of topologically massive
Yang-Mills theory in the Hamiltonian picture.  The material here is
standard and can be found in many places in the literature.  Here
we follow the original work of Deser, Jackiw and Templeton
~\cite{djt}.  In Section 3 we study the strong coupling limit of
that model and in Section 4 we show how it is related to $G/G$ models
and compute some of the Polyakov loop correlators.  In Section 5 we
examine the strong coupling model at finite temperature and find
that it is given by a perturbed G/G model.  In Section 6 we discuss
the quantum structure of the perturbed G/G model and show that the 
effective potential for the gauge group variables can be made finite 
at one loop.  In Section 7 we examine 
the nature of excited states of the strong coupling theory.  We construct
non-Abelian Landau levels on punctured
Riemann surfaces and show that these states saturate the thermal
ensemble of the theory. In Section 8 we summarize the conclusions.


\section{Topologically massive gauge theory revisited}

Consider a gauge theory in 2+1 spacetime dimensions where the basic
dynamical object is the Lie algebra-valued 1-form $A=A_\mu(x) dx^\mu$
whose density $A_\mu(x)$ is the gauge field.  The model has gauge
symmetry under $ A^g= g^{-1}Ag+ig^{-1}dg $ where $g(x)$ is a smooth
function on 2+1-dimensional spacetime taking values in the fundamental
representation of a compact semisimple Lie group. The action is
$S=S_{CS}+S_{YM}$, with $S_{CS}$ and $S_{YM}$ defined in
Eqs. (\ref{CS}) and (\ref{YM}) respectively.\footnote{{\it
Conventions}: The covariant derivative is $ D_\mu=\partial_\mu-iA_\mu
$ and the curvature tensor is $ F_{\mu\nu}=\partial_\mu
A_\nu-\partial_\nu A_\mu-i\left[ A_\mu,A_\nu\right] $.  Hermitean
generators of the Lie algebra $ \left[ T^a,T^b\right]=if^{abc}T^c $
are normalized in the fundamental representation as $ {\rm Tr} (T^a
T^b)=\frac{1}{2}\delta^{ab} $.  The connection and curvature are $
A_\mu=A_\mu^a T^a~~,~~F_{\mu\nu}=F_{\mu\nu}^a T^a $.}

We begin by reviewing the canonical quantization of topologically
massive gauge theory.  We suppose that all fields vanish
sufficiently fast at spatial infinity, so that the three-manifold we are
quantizing on is of the form $\Sigma\otimes\RR$, with $\Sigma$ the
Riemann sphere. The dynamical variables in phase space are the
spatial components of the gauge field, $A_i(\vec x)$, and the
associated canonical momenta, $\Pi_i(\vec x)$.  The equation of motion
for the temporal component of the gauge field $A_0$, Gauss' law
\begin{equation}
{\cal G}\equiv \frac{1}{e^2}D_i F_{0i}+\frac{k}{4\pi}F_{12} = 0 \ ,
\mabel{gauss}
\end{equation}
will be a constraint on the phase space variables.  The canonical
momentum conjugate to the spatial component of the vector field is
\begin{equation}
\Pi_i^a(\vec x)=\frac{\partial}{\partial \dot A_i^a(\vec x)}L
=\frac{1}{e^2}F_{0i}^a(\vec x) +\frac{k}{8\pi}\varepsilon_{0ij}A_j^a(\vec
x)
\end{equation}
We define the (unconventionally normalized) electric field as
\begin{equation}
E_i(\vec x)=\frac{1}{e^2}F_{0i}(\vec x)=\Pi_i(\vec x)-
\frac{k}{8\pi} \varepsilon_{ij} A_j(\vec x)
\mabel{elec}
\end{equation}
and the magnetic field as
\begin{equation}
B(\vec x)= F_{12}(\vec x)
\end{equation}

The Hamiltonian is
\begin{equation}
H=\int d^2x\left( e^2~{\rm Tr}\left( {\vec E}^2(\vec x)\right)
+\frac{1}{e^2}~{\rm Tr}\left(B^2(\vec x)\right)\right)
\end{equation}
The canonical commutation relation for the gauge field and its
canonical momentum is
\begin{equation}
\left[ A_i^a(\vec x), \Pi_j^b(\vec y)\right] =i\delta_{ij}\delta^{ab}
\delta(\vec x-\vec y)
\end{equation}

The electric field defined in (\ref{elec}) is analogous to the
velocity operator and differs from the canonical momentum by
${\cal A}^a_i(x;A) = k
\varepsilon_{ij} A_j^a(\vec x)/8\pi$.  The commutator of the electric
fields is
\begin{equation}
\left[ E_i^a(\vec x), E_j^b(\vec y)\right] =
-\frac{ik}{4\pi}\varepsilon_{ij} \delta^{ab}\delta(\vec x - \vec y )
\end{equation}
Also
\begin{equation}
\left[ A_i^a(\vec x), E_j^b(\vec y) \right] =i\delta^{ab}\delta_{ij}
\delta( \vec x - \vec y)
\end{equation}

In terms of phase space variables, the gauge constraint (\ref{gauss}) is
\begin{equation}
{\cal G}(\vec x)=
\partial_i  E_i(\vec x)-i\left[A_i(\vec x), E_i(\vec x) \right]
+\frac{k}{4\pi}B(\vec x)\sim 0
\mabel{gl}
\end{equation}
The weak equality, $\sim$, in (\ref{gl}) implies that, rather than
solving (\ref{gl}) as an operator equation, we will impose it as a
physical state condition in the quantum theory.  States in the physical
sector of the quantum Hilbert space are in the kernel of the constraint
operator
\begin{equation}
{\cal G}(\vec x)~\Psi_{\rm phys.}=0
\end{equation}
The operator ${\cal G}(\vec x)$ is the generator of infinitesimal
time-independent gauge transformations and commutes with gauge
invariant operators, such as the  energy density and the Hamiltonian,
\begin{equation}
\left[ {\cal G}(\vec x), ~H\right]=0 \ .
\end{equation}

In the functional Schr\"odinger picture, the gauge field operator
$\vec A(\vec x)$ is treated as a position variable.  The quantum
states are wave functionals of gauge field configurations, $\Psi[A]$,
and the canonical momentum is the functional derivative operator,
\begin{equation}
\Pi_i^a(\vec x)\equiv \frac{1}{i}\frac{\delta}{\delta A_i^a(\vec x)}
\end{equation}
The stationary states satisfy the functional Schr\"odinger equation
\begin{equation}
\int d^2x  \left( \frac{e^2}{2}
\sum_{i,a}
\left(\frac{1}{i}\frac{\delta}{\delta A_i^a(\vec x)}
- \frac{k}{8\pi}\varepsilon_{ij} A_j^a(\vec x)\right)^2
+\frac{1}{2e^2}\sum_a B^a(\vec x)^2
\right)\Psi[A]~=~{\cal E}~\Psi[A]
\end{equation}
where ${\cal E}$ is the energy of the state.  Among the stationary states,
the physical states are those which are in the kernel of the gauge
generator,
\begin{equation}
\left(
 D_i^{ab}(\vec x)
\left(\frac{1}{i}\frac{\delta}{\delta A^b_i(\vec x)}-
\frac{k}{8\pi}\varepsilon_{ij}A_j^b(\vec x)
 \right)
+\frac{k}{4\pi}B^a(x)\right)~\Psi_{\rm phys~}[A]=0
 \mabel{1gl}
\end{equation}

 Since the gauge constraint commutes with the Hamiltonian, the gauge
condition (\ref{1gl}) can be consistently imposed on eigenstates of the
Hamiltonian.  The presence of the magnetic field term in ${\cal
G}(\vec x)$ implies that the gauge symmetry is represented
projectively.  The invariance of the physical states under infinitesimal gauge
transformations in (\ref{1gl}) can be integrated.
Accordingly, the wavefunctional of a physical state obeys
\begin{equation}
\Psi_{\rm phys~}[A]~=~e^{i\alpha[A,g]}~\Psi_{\rm phys}[A^g]
 \mabel{projective}
\end{equation}
where the projective phase is given by the cocycle
\begin{equation}
\alpha[A,g]=\frac{-ik}{4\pi}
 \int d^2x {\rm Tr}(A~dgg^{\dagger})-\frac{k}{4\pi}\Gamma[g]
\ ,
\end{equation}
 related to the two-dimensional chiral anomaly
(see. e.g. \cite{fadsha,gr}). The last term in this equation
is the Wess-Zumino term \cite{wz}, written in terms of the
functional
 \be
\Gamma[g] = \frac13 \int_B \left(g^\dagger dg\right)^3
 \mabel{wz}
 \ee
 of the extension of the gauge transformation group element $g$ to the
3-ball $B$, which has the space as its boundary.  This term is
well-defined when the group valued field, $g(\vec x)$, has boundary
conditions such that the space can be considered as one-point
compactified.  This is sufficient for the two dimensional space to be
the boundary of a three-dimensional space.  The Wess-Zumino term is
well-defined up to an integer which does not appear in the
transformation law for the wavefunctional if the coefficient $k$ is
quantized.  This requirement of finding a single-valued solution to the
gauge constraint is the reason for quantization of the coefficient of
the Chern Simons term in the Hamiltonian formalism.


\section{Strong coupling limit}

The strong coupling limit, $e^2\rightarrow\infty$, of the
topologically massive gauge theory is obtained in the Hamiltonian
formalism by neglecting the magnetic term in the Hamiltonian.
Physically, this is the limit of large topological mass, where all
momenta of interest are less than the topological mass
$\mu=e^2k/4\pi$.  In this limit, since the gluon has infinite mass, we
expect that the model has no degrees of freedom.

Since the strong coupling theory is not generally
renormalizable, it is usually considered in the context of
lattice gauge theory ~\cite{wilson}. In 2+1 dimensions, this
is particularly obvious, since the coupling constant $e^2$
has dimensions of mass.  Therefore, the large coupling
limit, particularly of quantities with no momentum or
coordinate dependence, can only be taken relative to some
scale, such as the ultraviolet momentum cutoff,
$\Lambda^2$. In a later section we shall make some
conjectures about how the coupling constant can be adjusted
as a function of the cutoff so that some thermodynamic
quantities turn out finite.

The strong coupling Hamiltonian is
\begin{equation}
H_0=e^2\int d^2x ~{\rm Tr}\left({\vec E}^2(\vec x)\right)
\mabel{h0}
\end{equation}
The energy densities at different points commute,
\begin{equation}
\left[ ~{\rm Tr}\left({\vec E}^2(\vec x)\right)
~, ~{\rm Tr}\left({\vec E}^2(\vec y)\right)~
\right]=0
\end{equation}
and they also commute with ${\cal G}(\vec x)$.  Therefore, we can
search for simultaneous eigenfunctionals of these {\it densities}.

Let us assume that $k$ is positive.  Consider the
holomorphic factorization of the energy density
\begin{equation}
~{\rm Tr}\left({\vec E}^2(\vec x)\right) ~=~~{\rm
Tr}\left(E^{\dagger}(\vec x)E(\vec x)\right)+
\frac{k}{4\pi}\delta^2(\vec 0)
\mabel{fact}
\end{equation}
where the ``creation'' and ``annihilation'' operators are
\begin{equation}
E(\vec x)= E_1(\vec x)-iE_2(\vec x) ~~,~~
E^{\dagger}(\vec x)=E_1(\vec x)+iE_2(\vec x) ~~,
\mabel{crea-ann}
\end{equation}
and the divergent term in Eq. (\ref{fact}) is the vacuum energy density
arising from normal ordering the energy density using the commutator
\begin{equation}
\left[ E^a(\vec x), {E^b}^{\dagger}
(\vec y) \right] =  \frac{k}{2\pi} \delta^{ab}
\delta( \vec x - \vec y)
\end{equation}
Corresponding to the operators $E,E^\dagger$, we have the conjugate
anti-holomorphic and holomorphic fields $A_\pm \equiv A_1\pm iA_2$,
with the non-vanishing commutators
\be
 \left[ A_+^a(\vec x), ~E^b(\vec y) \right] =
 \left[ A_-^a(\vec x), ~E^{b\dagger}(\vec y) \right] =
 2 i ~\delta^{ab}\delta(\vec x-\vec y)
 \ee

As usual, we define the normal ordered energy density by dropping the
divergent term:
\begin{equation}
:{\rm Tr}\left( E^2(\vec x)\right):~~\equiv~
{\rm Tr}\left( E^{\dagger}(\vec x) E(\vec x) \right)
\mabel{normalo}
\end{equation}
The vacuum state is the zero mode of the annihilation operator,
\begin{equation}
E^a(\vec x)~\Psi_0[A]=
\left(
\frac{2}{i}\frac{\delta}{\delta A_+^a(\vec x)}
 -i\frac{k}{8\pi}A_-^a(\vec x)\right)\Psi_0[A]=0
\end{equation}
If normalizable solutions of this equation exist, then they can be
superposed to form eigenstates of ${\cal G}(\vec x)$.  Those
annihilated by ${\cal G}(\vec x)$ are gauge invariant ground states.

The vacuum-equation is solved by any functional of the form
\begin{equation}
\Psi_0[A]=\psi[ A_-]~\exp\left\lbrace -\frac{k}{8\pi}\int d^2x~
{\rm Tr}\left(\vec A^2 (\vec x)\right)\right\rbrace
\mabel{vac}
\end{equation}
where $\psi$ is a holomorphic wave-functional, depending only on $A_-$
of the two components $A_\pm$.  This solution is degenerate in that any
functional $\psi$ is acceptable, subject to the normalizability of the
wave-functional, implemented by functional integration
\begin{equation}
1=\int ~[dA] ~\Psi_0^{\dagger}[A]~\Psi_0[A]
\end{equation}
which is consistent when $k>0$.  The degeneracy is resolved by
requiring gauge invariance,
\begin{equation}
{\cal G}(\vec x)\Psi_0[A]=0
\end{equation}
This yields the equation
\begin{equation}
\left(
 D_-^{ab}(\vec x) \frac{\delta}{\delta A_-^b(\vec x)} -
\frac{k}{8\pi}\partial_+A_-^a(\vec x)
 \right)\psi[A_-]=0
 \mabel{gaug}
\end{equation}
where $\partial_\pm=\partial/\partial x_1\pm i \partial/\partial x_2$.
This is the anomaly equation for $k$ flavors of chiral fermions in
two-dimensional Euclidean space\footnote{The holomorphicity
(\ref{vac}) and gauge-invariance (\ref{gaug}) conditions on the vacuum
wave-functional are familiar from the Bargmann quantization of
Chern-Simons theory \cite{witt3,dunne,emss}, where the space of
spatial gauge fields on a Riemann surface, subjected to a gauge
constraint generated by a flatness condition, is quantized with respect
to the K\"ahler potential $\int A_+ A_-$.  The appearance of these
conditions just rephrases the fact that Chern-Simons theory
describes the vacuum sector of strongly coupled gauge theory.}.  A
quantity that satisfies this equation is the $k$'th power of the
determinant of the chiral Dirac operator in two dimensions
\begin{equation}
\psi[A_+]~=~{\rm const.}~\left[\det D_+ \right]^k
 \mabel{chirdet}
\end{equation}
where the chiral Dirac operators,
in a Weyl basis,  are
\begin{equation}
D_+~=~\left( \matrix{  0 & \partial_-\cr -\partial_++iA_+&0\cr}\right)
~~,~~
D_-~=~\left( \matrix{  0 & \partial_--iA_-\cr -\partial_+&0\cr}\right)
\end{equation}
The normalization integral for the wave-function is
\begin{equation}
Z=\int [dA]~\left[\det D_- \right]^k
~\left[\det D_+ \right]^k
\exp\left\lbrace-\frac{k}{4\pi}\int d^2x {\rm Tr}\left(
{\vec A}^2(\vec x)\right)\right\rbrace
 \mabel{holfac}
\end{equation}
Note that the terms in the exponent are exactly the counterterms which
are necessary to convert the holomorphically factorized determinant of
the Dirac operator into the gauge invariantly regulated determinant of
the vector-coupled Dirac operator (see e.g. \cite{almova}),
\begin{eqnarray}
Z&=&\int [dA]~
~\left[\det \slash D\right]^k \mabel{detdet}
\\
&=&\lim_{g^2\rightarrow\infty}
\int [dA][d\psi d\bar\psi] e^{-\int d^2x\left( \sum_{\alpha=1}^k
\bar\psi_\alpha \slash D \psi_\alpha+\frac{1}{2g^2}{\rm Tr}F_{\mu\nu}^2\right)}
\mabel{sc}
\end{eqnarray}
where we have absorbed a factor of $~\det^k(-\partial^2)$ into the
normalization.  In both (\ref{detdet}) and (\ref{sc}),
the Dirac determinant is assumed to have a gauge invariant
regularization.
Eq. (\ref{sc}) is the strong coupling limit of two
dimensional QCD with $k$ flavors of massless quarks.  Furthermore, the
computation of any observable functional of $\vec A(\vec x)$ is
equivalent to the computation of the equivalent expectation value in
the two-dimensional Euclidean gauge theory
\begin{equation}
\langle 0\vert {\cal O}[A]
\vert0\rangle=\lim_{g^2\rightarrow\infty}\frac{1}{Z} \int [dA][d\psi
d\bar\psi] ~{\cal O}[A] e^{-\int d^2x\left( \sum_{\alpha=1}^k
\bar\psi_\alpha \slash D \psi_\alpha +\frac{1}{2g^2}{\rm
Tr}F_{\mu\nu}^2\right)} \mabel{exp}
\end{equation}
For example, the vacuum expectation value of the spatial Wilson loop
in 2+1-dimensional gauge theory is the expectation value of a
spacetime Wilson loop in the infinite coupling limit of two-dimensional
QCD.  In the latter theory, since the quarks transform under the
fundamental representation of the gauge group, the Wilson loop
exhibits a perimeter law,
\begin{equation}
\left\langle {\rm Tr}\left( {\cal P}e^{i\oint_\Gamma A}\right)\right\rangle
\sim \exp\left\lbrace -\sigma L[\Gamma]\right\rbrace
\end{equation}
This is a signal of the magnetic mass of the topological field theory
and is consistent with the fact that topologically massive gauge
theory is not confining.


\section{Bosonization, Polyakov loop correlators and the Verlinde
 formula \label{BPV}}

To facilitate computations, we bosonize the functional
integral (\ref{exp}) using non-Abelian bosonization
\cite{witt4}. Two-dimensional fermion determinants in a
background field $A$ were computed by Polyakov and
Wiegmann \cite{pw}, resulting in a principal chiral model
with a Wess-Zumino term. The corresponding functional is
the celebrated Wess-Zumino-Novikov-Witten action
\cite{novi,witt4},
\be
I[h] = \frac{1}{8\pi} \int d^2x~\Tr\left| \nabla h \right|^2 +
\frac{i}{4\pi} \Gamma[h] \ ,
 \mabel{wzw}
\ee
 with $\Gamma$ the Wess-Zumino functional of
Eq. (\ref{wz}).  The group valued field $h$ is an
hermitian combination $h=\tilde h \tilde h^\dagger$ of a
field $\tilde h$ taking values in the complexification
${\rm G}_\CC$ of the gauge group, which describes the
background field as
\be
A_+=i \tilde h^{- 1}\del_+ \tilde h \ ,~~~~
A_-=i \del_- \tilde h^\dagger(\tilde h^\dagger)^{- 1} \ .
\mabel{Apm}
\ee

Using the Polyakov-Wiegmann formula,
\be
I[\tilde h \tilde h^\dagger] = I[\tilde h] + I[\tilde
h^\dagger] - \frac{1}{4\pi} \int d^2x ~~\tilde h^{- 1}\del_+
\tilde h ~~\del_- \tilde h^\dagger(\tilde h^\dagger)^{- 1} \ ,
 \mabel{pwf}
\ee
we get the fermion determinant exactly in the
holomorphically factorized form corresponding to
Eq. (\ref{holfac}).

The WZNW functional describing the fermion determinant is
invariant under vector gauge transformations, $\tilde h
\to \tilde h g\ , ~ \tilde h^\dagger \to g^\dagger \tilde
h^\dagger$. However, under a chiral gauge transformation,
$\tilde h \to \tilde h g \ , ~\tilde h^\dagger \to g \tilde
h^\dagger$, it transforms non-trivially. Using this
property, we can write an equivalent bosonic action for
the fermionic functional integral, in terms of the group
valued field $g\in{\rm G}$ \cite{divecc}. The resulting
theory is a gauged WZNW theory \cite{gk},
more exactly the so called G/G WZNW theory with the action
\be
 S_{{\rm G/G}}[g,A] = I[g] + \frac{1}{4\pi} \int d^2x \Tr\left(i
A_-\partial_+ g g^{-1} + A_+ A_- - A_-^g A_+ \right) \ .
 \mabel{GWZW}
\ee
 Here we have chosen to use the  gauge fields instead of
the  group valued field $\tilde h$.

For a generic topological mass, we are dealing with the
$k$:th power of the fermion determinant, so the required
bosonization is somewhat more complicated \cite{fs}. In
addition to the gauged WZNW model for the color degrees of
freedom, now at level $k$, there are ungauged ones for the
flavor degrees of freedom
and for a field parametrizing the coset of
the total ungauged symmetry group of the fermions divided
by the gauge and flavor subgroups of the total symmetry group.
Here we shall only be
interested in the color degrees of freedom, which decouple
from the rest as there is no mass term for the would-be
bosonized fermions.  The normalization integral for the
vacuum wave functional (\ref{sc}) thus becomes
\be
 Z = \int [dA][dg] \e^{-kS_{{\rm G/G}}} \ ,
\ee
 which is the partition function for the
 G/G model at level $k$.

The result that the vacuum sector of strongly coupled topologically
massive gauge theory is described by a G/G WZNW model is by no means
surprising. The connections of pure Chern-Simons theory, which
describes the vacuum sector, with G/G models have been widely
investigated in the literature \cite{spiegel,witt1,bt1,geras}.  In
particular, in Refs. \cite{bt1,geras} it was shown that the finite
temperature effective action for Chern-Simons theory, defined on the
three-manifold $\Sigma\otimes S^1$ with $\Sigma$ an arbitrary Riemann
surface, is nothing but the G/G action. There is no temperature
dependence, as the Hamiltonian of pure Chern-Simons theory is zero.

Moreover, the properly normalized partition function of G/G WZNW on a
Riemann surface $\Sigma$ is an integer, characterizing the dimension
of the Chern-Simons Hilbert space when quantized on
$\Sigma\otimes\RR$. These integers are given by the so-called Verlinde
formula \cite{verlinde}, where they appeared as the dimensions
of the spaces of conformal blocks of a WZNW model for the group G. We
shall be interested in the simplest case when $\Sigma$ is a sphere,
possibly with punctures.

The G/G partition function on a sphere is just 1, confirming the
uniqueness of our vacuum (\ref{vac}). Punctures, on the other hand,
are labeled by representations $\lambda_i$ of the group. Most
interesting for us is the observation \cite{witt3,bt1} that punctures
on $\Sigma$ correspond to traces of vertical Wilson loops in
representations $\lambda_i$ when quantizing Chern-Simons theory on
$\Sigma\otimes S^1$, ~i.e. nothing but the Polyakov loops of finite
temperature gauge theory \cite{poly,suss}.

Polyakov loops in Chern-Simons theory correspond in G/G to traces of
the group-valued field $g$, i.e. characters
$\X_{\lambda_i}(x_i)\equiv\Tr \bigl[g(x_i)\bigr]_{\lambda_i}$ in the
representations $\lambda_i$.

The correlation functions
\be
 \left< \prod_i \X_{\lambda_i}(x_i) \right> = \int[dA][dg] \prod_i
\X_{\lambda_i}\left(x_i\right) \e^{-kS_{{\rm G/G}}[g,A]}
\ee
are given by the Verlinde formula for the dimensions of the space of
conformal blocks on the corresponding punctured surfaces.
Specializing to the case of gauge group G=SU(2), and integrable
($\lambda\leq k$) representations of the corresponding Ka\v c-Moody
group, we have the lowest-point functions (see e.g. \cite{bt1})
\be
  ~~~~~~~~~\left<  \X_\lambda(x) \right> = 0\ ; ~~\lambda>0
 \mabel{polyexp}
\ee
\be
  \left< \X_\lambda(x) \X_\nu(y)  \right> =
\delta_{\lambda\nu} \mabel{polycorr}
\ee
\be
  \left<  \X_\lambda(x)\X_\mu(y)\X_\nu(z) \right> =
N_{\lambda\mu\nu}
\mabel{ver3}
\ee
 Here $N_{\lambda\mu\nu} $ is the matrix encoding the fusion rules
of conformal blocks corresponding to different primary fields inserted at
the punctures.

As explained above, this enables us to calculate correlators of
Polyakov loops in any representation.  As G/G WZNW theory is a
topological field theory, the correlators are just topological
invariants pertaining to the corresponding punctured spheres.  These
correlators can then be interpreted as zero temperature correlators in
strong coupling topologically massive gauge theory.  For pure
Chern-Simons theory, corresponding to infinite coupling, $e^2=\infty$,
expressions (\ref{polyexp})-(\ref{ver3}) give the Polyakov loop
correlators in the full finite temperature theory. The latter is of
course rather trivial, as in a topological theory there are no states
at finite energy and therefore no excited states to give a temperature
dependence to the partition function or correlators.

Polyakov loops correspond to infinitely massive external particle
insertions, and their correlators are of vital importance when
investigating the confinement properties of gauge theories in the
absence of dynamical fundamental fermions \cite{poly,suss}. This is
due to the fundamental representation Polyakov loop's ability to probe
the spontaneous breaking of the hidden symmetry of the theory under
global transformations in
the center of the gauge group (see Ref. \cite{sve} for a comprehensive
review).

Here one has to notice that the results for Polyakov loop correlators
quoted above are computed on a finite volume Riemann surface. Eq.
(\ref{polyexp}) indicates that the expectation value of a Polyakov
loop is always vanishing in finite volume. On the other hand, cluster
decomposition of the correlator (\ref{polycorr}) with fundamental
representation Polyakov and anti-Polyakov loops gives a {\it
non-vanishing} expectation value for the Polyakov loop in the infinite
volume limit.  This indicates spontaneous breaking of the center
symmetry in infinite volume.  According to \cite{poly,suss}, this symmetry
breaking indicates the de-confinement of color in the theory.  This is
commensurate with what we would expect for this model, since the gluons are
very massive, and therefore incapable of mediating a confining interaction.

For topologically massive gauge theory,
eqs. (\ref{polyexp})-(\ref{ver3}) give vacuum correlators of Polyakov
loops. To investigate the full thermodynamical expectation values, one
has to find the effective thermodynamical theory for
$e^2\neq\infty$. This will be done in the next section. However, there
is no reason to expect that the long distance correlations expressed
in (\ref{polycorr}) would vanish by exiting some local quantum
states. Accordingly, we expect the center symmetry to be broken in the
topologically massive theory as well, which is consistent with the
fact that the latter theory is not confining.

The screening and confinement properties of the Abelian topologically
massive theory (at finite temperature and any coupling) were
investigated in Ref. \cite{gsst2}. There it was noticed that the
effective action is of the same Villain form as the one obtained for
the Schwinger model in \cite{gsst1}, although one dimension
higher. Detailed calculations in \cite{gsst1} show an analogous
picture as the one emerging here from the analysis of
Eqs. (\ref{polyexp}) and (\ref{polycorr}) in the $e^2=\infty$ case: At
finite volume, the Polyakov loop expectation value is always zero.  In
the infinite volume limit one has to rely on cluster decomposition of
the correlator of two Polyakov loops.
The center symmetry is broken and the theory
is in a deconfined phase.\footnote{In the Abelian theories, the
analogue of the center symmetry is the $Z$ symmetry of the finite
temperature theory, and the order parameter transforming under this
symmetry, corresponding to an fundamental Polyakov loop, is the
Polyakov loop of a particle with an incommensurate charge compared to
the basic charge of the theory.}


\section{Finite temperature theory  \label{FTT}}

Let us now consider the strongly coupled theory at finite temperature.
The partition function is the trace of the Boltzman factor over the
physical states.
We will use the density matrix
\begin{equation}
\rho~=~ e^{-H/T}\cdot{\cal P}
\end{equation}
where $H$ is the Hamiltonian and ${\cal P}$ is the projection operator onto
physical states.
To choose physical states, we shall use position
eigenstates which have the property
\begin{equation}
\hat {\vec A}(\vec x)~\vert A\rangle ~=~\vec A(\vec x)~\vert A\rangle
\end{equation}
We project onto gauge invariant states by integrating over all gauge
copies of the state
\begin{equation}
{\cal P}\left| A\right>~=~\frac{1}{\rm vol~{\cal G}}
\int [dg] ~e^{i\alpha[A,g]}~\left| A^g \right>
\end{equation}
where $[dg]$ is a local Haar measure and ${\rm vol~{\cal G}}$ is the volume of
the gauge group.  Using this, the partition function is (up to an overall
constant)
\begin{equation}
Z[T]=\int[dg]dA~ e^{i\alpha[A,g]}\langle A\vert~
e^{-H/T}~\vert A^g\rangle
\end{equation}

We shall consider the strong coupling limit where $H\rightarrow H_0$.  This
limit could be considered as the limit $e^2T\rightarrow \infty$ (again since
this is a dimensionless number, it should be read as infinity compared to all
other mass or momentum scales of interest) while keeping the dimensionless
ration $e^2/T$ constant.
We will compute the propagator
$$
\langle A\vert~ e^{-H_0/T}~\vert A^g\rangle ~.
$$
The propagator satisfies the Heat equation,
\begin{equation}
\left[\frac{\partial}{\partial T^{-1}}+ H_0 \right]~
\langle A\vert~e^{-H_0/T}~\vert A'\rangle~=~0 ~,
\end{equation}
with the boundary condition
\begin{equation}
\lim_{T\rightarrow\infty}\langle A\vert~e^{-H_0/T}~\vert A'\rangle
=\prod_{x,a,i}\delta( A_i^a(\vec x)-A_i^a(\vec x)') ~.
\end{equation}
This equation is solved by the Gaussian functional
\begin{equation}
\langle A\vert e^{-H_0/T}\vert A'\rangle =
\frac{\exp{\left\lbrace -\frac{k}{8\pi}\int d^2x~~{\rm
Tr}\left( \coth\left(\frac{ke^2}{8\pi T}\right) \left(\vec A-\vec
A'\right)^2 + 2i\vec A\times \vec A' \right)
\right\rbrace}}{\left(\frac{16\pi^2}{k} \sinh \left(
\frac{e^2 k}{8\pi T}\right)\right)^{({\rm dim}~G)~V~\delta(0)}}
\end{equation}
 where $V$ is the volume of the two-manifold $\Sigma$.

Combining the propagator with the cocycle $e^{i\alpha[A,g]}$, the
result for the partition function is
\begin{equation}
Z[T]= 1/\N \int[dg(\vec x)][d\vec A(\vec x)]~e^{-S_{\rm eff}[A,g]} ~,
\mabel{ftpart}
\end{equation}
where the finite temperature effective action is \footnote{The covariant
derivative acts on $g$ with the adjoint action $D_\mu
g(x)=\partial_\mu g(x)-i\left[ A_\mu(x),g(x)\right]$, so
that $g^\dagger D_\mu g$ equals the finite gauge
transformation $A_\mu^g - A_\mu$.}
\begin{eqnarray}
S_{\rm eff}[A,g] &=& \frac{k}{8\pi} \coth\left({\frac{k e^2}{8\pi T}}
\right) \int d^2x~{\rm Tr} \vert \vec D g\vert^2 + \frac{ik}{4\pi}
\Gamma[g,A] ~; \mabel{seff} \\ \Gamma[g,A] &=& \Gamma[g] + \Tr\int
\vec A \times \left(\vec A^g + i\vec\partial g g^\dagger
\right) \ ,\mabel{gamma}
\eea
 and the normalization is given by
\be
 \N = \left(\frac{16\pi^2}{k} \sinh \left(
\frac{e^2 k}{8\pi T}\right)\right)^{({\rm dim}~G))~V~\delta(0)} \ .
 \mabel{norm}
\ee
  This is the gauged principal chiral model with a gauged Wess-Zumino
term, i.e. a perturbed $G/G$ WZNW model.  In
the limit where $\coth(e^2 k/8\pi T)\rightarrow 1$, which is achieved
by $ke^2/T\rightarrow\infty$, it approaches the level
$k$ $G/G$-model.
The effective action (\ref{seff}-\ref{gamma}) can 
also be obtained starting from the 
propagator of a particle in a constant magnetic field \cite{glass} and 
extending the result to infinite degrees of freedom.
At each point the hamiltonian density in the strong coupling limit, 
is in fact the hamiltonian 
of a particle of mass  $1/e^2$ in a 
magnetic field $k/4\pi$.

Perturbations of $G/G$ WZNW models of the form (\ref{seff}) have been
earlier discussed by Witten \cite{witt2} and Blau \& Thompson
\cite{bt2}. In \cite{witt2} it was argued that the theory is
topological at the classical level, but the topological invariance is
broken due to quantum effects. In \cite{bt2}, a supersymmetrized
version of (\ref{seff}) was investigated. The supersymmetrization was
chosen so that the perturbed model was equivalent to the unperturbed
G/G model in the equivariant localization sense. Here, we shall
consider the truly perturbed G/G model, and find that the topological
invariance is indeed broken.


\section{One loop effective potential for $g$  \label{renor} }

Now we shall investigate the finite temperature effective field
theory built on the perturbed G/G action (\ref{seff}). We shall
mainly be interested in the effects of the perturbation $k\to k
\coth\frac{ke^2}{8\pi T}\quad$\footnote{It is a well known fact, 
both in G/G models 
\cite{bt1,geras,bt2}
and more generally in topologically massive gauge theory
\cite{pr,kmpp} that the level $k$ gets shifted by quantum effects
to $k+c_v$, where $c_v$ is the quadratic Casimir in the adjoint
representation. In this paper we are not interested in this issue.}.
  
By integrating out
the quantum fluctuations to one-loop order we shall find the
effective potential for the group valued field $g$.

We expand $g$ in terms of the constant background field
$g_o$ and algebra valued fluctuations $\sigma(x)$\footnote{Note that 
with a constant background the expansion in terms of the coordinates 
$\sigma$, coincides with the normal coordinate expansion.}; 
\[
 g = g_o~\e^{i \sigma} \ .
\]
 The piece of the finite temperature effective action (\ref{seff})
quadratic in the quantum fields is
\bea
S_q[g_o;\sigma,A_i] =  \frac{1}{4\pi} \int\Tr \Bigl\lbrace
 \frac{\kappa}{2} (\del_i\sigma)^2
 ~-~ 2 i k \varepsilon^{ij} A_i \del_j\sigma  ~~~ ~~~ ~~~ ~~~ ~~~
 \cr
 ~+~  \left(\unit - \Ad_{g_o}\right)(A_i)  \P^{ij} (A_j + \del_j\sigma)
 \Bigr\rbrace \ ,
\eea
 where $\Ad$ is the adjoint action of the group, $\P^{ij} = \kappa
\delta^{ij} + i k \varepsilon^{ij}$ is the deformed projector onto
holomorphic forms, and
\be
 \kappa = k \coth\left(\frac{k e^2}{8\pi T}\right)
\mabel{kappa}
\ee
 is the perturbed level.

In order to regulate the integration over the gauge field $A$, we add
the Yang-Mills term multiplied by an infinitesimal parameter,
$\epsilon\int {\rm Tr}(F^2)/8\pi$, to the action.  In the original
(2+1)-dimensional language, this regulator is the magnetic field
squared term which was killed by the strong coupling limit.  We shall
later take the limit where its coefficient $\epsilon$ goes to zero
like $\epsilon=1/\Lambda^2$.  We shall also add a covariant gauge
fixing term $\lambda~\int{\rm Tr} (\nabla\cdot A)^2/4\pi$.

Defining
\be
 R^{ab}(g) = 2 ~\Tr\left( (\unit - \Ad_g)(T^a) ~T^b\right)
	  = \delta ^{ab} - 2 ~\Tr\left(gT^a g^{-1} T^b\right) \ ,
 \mabel{rab}
\ee
  we can write the action in terms of the
coordinate fields in the  Lie-Algebra:
\bea
 S_q =  \frac{1}{8\pi} \int \left\lbrace
 - \frac{\kappa}{2} \sigma^a\Delta\sigma^a
 ~-~ \left(R^{ab}(g_o) \P^{ij} - 2 i k \delta^{ab}
\varepsilon^{ij}\right) \sigma^b \del_j A_i^a ~~~ ~~~ ~~~
 \right.\cr\left.
 ~+~  A_i^a R^{ab}(g_o) \P^{ij} A_j^b
 ~+~ \epsilon (\del_iA_j^a - \del_jA_i^a) \del_iA_j^a
 ~+~ {\lambda}\del_iA_i^a~\del_jA_j^a
\right\rbrace \ .
 \mabel{quad}
\eea
 The $\sigma$ field is now easily integrated out. In order
to facilitate the ensuing $A$-integration, we change
integration variables to the curl and divergence of
$A$. Using the identity
\[
 \delta^{ij}(x-y) ~=~ \varepsilon^{ik}\del_k \frac{1}{\Delta}
\varepsilon^{jl}\del_l ~+~ \del_i \frac{1}{\Delta}\del_j
\]
  we can write the term quadratic in $A$ in terms of $C^a =
\nabla\times A^a$ and $D^a= \nabla\cdot A^a$. The Jacobian
arising from the change $A_i^a \to \{C^a,D^a\}$ is just
proportional to the determinant of the Laplacian. After
integrating over $\sigma^a$, we have the effective action
\be
 S_{\tilde q} = \frac{1}{8\pi} \int \left\{
 \frac{1}{\kappa} \left( 2 k^2 \delta^{ab} + (\kappa^2 -
k^2) R^{(ab)} \right) C^a \frac{-1}{\Delta} C^b
 ~+~ {\epsilon} (C^a)^2
 ~+~ {\lambda} (D^a)^2 \right\} \ .
 \mabel{CDact}
\ee
 In deriving (\ref{CDact}) we used properties of $R$ that
follow directly from its relation (\ref{rab}) to the adjoint
representation matrix. The symmetrized matrix $R^{(ab)}$ is
Hermitean, so we can diagonalize it. We denote the
corresponding Eigenvalues $r^a$.

Next, we integrate out $D^a$ and $C^a$. This gives the one-loop
contribution to the partition function
\[
 Z_q \sim \prod_a \left[\Det\left( \frac{2 k^2 - \epsilon \kappa
\Delta}{-\kappa\Delta} \right) \Det\left(1 + \frac{
(\kappa^2-k^2) ~r^a(g_o)}{2 k^2 - \epsilon\kappa\Delta}
 \right)\right]^{-1/2} \ .
\]
 From this, we can read out the effective potential for
$g_o$:
\be
 V_{\rm eff}[g_o] = \frac{V}{2} \int \frac{d^2p}{4\pi^2}
   ~\ln\left[ 1 ~+~ \frac{ (\kappa^2-k^2) ~\Tr~R(g_o)}{2 k^2
+ \epsilon\kappa p^2} + \O\left((\kappa^2-k^2)^2\right)
 \right] \ .
 \mabel{veff}
\ee
 The matrix $R$ is traced over adjoint indices, $Tr~R=N^2-|\Tr~g|^2$. The
expansion is around the zero temperature (infinite
coupling) conformal fixed point with $\kappa = k$, which
corresponds to the $G/G$-model.

With a fundamental cut-off $\Lambda$, we can perform the integration
over $p$. In the vicinity of the conformal fixed point we thus have
\be
 V_{\rm eff} \sim
 \Lambda^2 \ln\left(1 ~+~\frac{a}{b+\epsilon\Lambda^2}
\right)
~+~ \frac{a+b}{\epsilon} \ln\left(1
~+~ \frac{\epsilon\Lambda^2}{a+b} \right)
 ~-~ \frac{b}{\epsilon} \ln\left(
\frac{b+\epsilon\Lambda^2}{b} \right) \ .
 \mabel{renorski}
\ee
 Here we have denoted $a=\Tr~R(g_o) ~(\kappa^2-k^2)/\kappa$ and
$b=2k^2/\kappa$. This indicates that the effective potential is finite
if the regulator is taken to zero according to $ \epsilon \sim
\Lambda^{-2}$. In addition, we have to require that
$a\sim Tr~R m^2/\Lambda^2$, for some scale $\mu$. Recalling the definition 
(\ref{kappa}) of $\kappa$,
this means that $e^2/T$ has to scale logarithmically with the cutoff.

Our effective theory (\ref{seff}) for the thermodynamics of
strongly coupled topologically massive gauge theory thus
makes sense as long as we stay within a logarithmic
distance from the cutoff, i.e. within a coupling-to-temperature range
$e^2/T >\sim \ln\Lambda/\mu$.

The effective potential (\ref{renorski}) has absolute minima for all the 
gauge group elements belonging to the center of the gauge group, $g_0\in 
Z_N$. For such a $g_0$, $\Tr~R(g_0)=a=0$, and the one-loop 
effective potential vanishes. Consequently, the Polyakov loop operators 
tend to be elements of $Z_N$ so that the center symmetry is broken.
This leads to a non-vanishing vacuum expectation value for a Polyakov loop 
operator, consistently with the discussion of Section~\ref{BPV} on the 
correlators of Polyakov loop correlators. 
 We can then conclude that in the strong 
coupling limit topologically massive 
gauge theory is always in a deconfined phase, as one would expect 
in this regime where the gluons have a big mass.


\section{Excited states}

In this Section, we shall find the excited states of the
theory,  which give rise to the finite
temperature partition function (\ref{ftpart}) in the
regime for the coupling constants, discussed in the previous Section.

The natural candidate for excited states are the Landau levels created
by the field theoretic harmonic oscillator creation operator
$E^\dagger$ of Eq. (\ref{crea-ann}), acting on the vacuum
(\ref{vac}). The electric field transforms covariantly under a gauge
transformation, so some care has to be taken to render the emerging
states gauge invariant up to the projective phase of
Eq. (\ref{projective}).  We shall see that this will be possible by
employing the Chern-Simons Hilbert spaces on surfaces with punctures
which were discussed in Section \ref{BPV}.

Using the results of Ref. \cite{pw} cited in
Eqs. (\ref{wzw}---\ref{pwf}), we would straight forwardly get the
holomorphically factorized ground state. To work within this
formulation, we would then have to rewrite the electric fields (the
gauge field functional derivatives) in terms of cotangent operators on
the complex group manifold parametrized by $\tilde h$. This approach
was taken in Refs. \cite{karnai}, where the related problem of the
origin of the mass gap in 2+1 dimensional gauge theories was
addressed.

For our purposes, it will be most transparent to avoid the inherent
nonlinearities associated with quantizing the curved group manifold,
and work on the level of the Lie-algebra instead. This can be achieved
in the gauged WZNW formulation.  The price one has to pay for working
with a vector space formulation of the gauge fields, is increased
complications in the holomorphic factorization. The inclusion of the
bosonic field $g$ spoils the straight forward factorizability given by
the Polyakov-Wiegmann formula (\ref{pwf}).

In Ref. \cite{witt1} Witten showed that gauged WZNW models can be
factorized only if one introduces an extra gauge field $B$ (see also
\cite{bht}). By introducing this field, and using the
Polyakov-Wiegmann formula, as well as the invariance of the Haar
measure, it is easy to show that the gauged WZNW partition function can
be factorized as
 \be
 Z = \int [dA][dg] \e^{-kS_{{\rm G/G}}} = {\textstyle \frac{1}{{\rm
vol}~{\cal G}}} \int [dA][dB] ~\Psi^*_0[A,B] ~\Psi_0[A,B] \ ,
 \mabel{factAB}
 \ee
 where the vacuum wave functional depending on
the two gauge fields is
\be
 \Psi_0[A,B] = \e^{-\frac{k}{8\pi}\int \left(\vert A\vert^2 + \vert
B\vert^2 \right) } ~\int [dg] \e^{-I[g] + \frac{k}{4\pi}
\int\left\{A_- B_+^g - i B_+ \del g g^{-1} \right\}} \ .
\ee
  The extra gauge field $B$ records the fact that the group
$G$ in the gauged WZNW model is diagonally gauged. This
translates into the symmetry of $\Psi_0[A,B]$ under an
exchange of $A$ and $B$ and complex conjugation:
 \be
  \Psi^*_0[A,B] =  \Psi_0[B,A]  \ .
 \mabel{xchange}
\ee

Under a gauge transformation, $\Psi_0[A,B]$ transforms
as
 \be
 \Psi_0[A^h,B^l] = \e^{-i \alpha[A,h] ~+~ i \alpha[B,l]}
~\Psi_0[A,B] \ ,
\ee
 generalizing the projective transformation of physical
states (\ref{projective}) to two gauge
fields. Similarly, it is easy to see that $\Psi_0[A,B]$
is annihilated by the annihilator $E$:
\be
 E^a~\Psi_0[A,B]=
\left(
\frac{2}{i}\frac{\delta}{\delta A_+^a}
 -i\frac{k}{8\pi}A_-^a\right)\Psi_0[A,B]=0\ .
\ee
 The wave functional $\Psi_0[A,B]$ thus fulfills the
requirements for a vacuum wave functional for the
strongly coupled massive gauge theory. Moreover, due to
Symmetry (\ref{xchange}) the analogous operator in $B$
acts as an annihilator as well:
 \be
 \left( \frac{2}{i}\frac{\delta}{\delta B_-^a}
-i\frac{k}{8\pi}B_+^a\right)\Psi_0[A,B]=0\ .
\ee

\bigskip

In order to build gauge-invariant excited states, we take a look at
Chern-Simons Hilbert spaces on punctured Riemann spheres. As explained
in Section \ref{BPV}, punctures $x_i$ correspond to Polyakov loop
insertions, which again correspond to non-dynamical external
particles. The particles, and accordingly the punctures, are
characterized by the representations $\lambda_i$ of the group $G$,
under which they transform. The corresponding wave functionals can be
acquired by inserting the group element $g$ in representations
$\lambda_i$ at the points $x_i$, into the vacuum functional
$\Psi_0[A,B]$ without punctures:
 \be
  \Xi_{\{\lambda_i\}}[\{x_i\};A,B] =
\e^{-\frac{k}{8\pi}\int \left(\vert A\vert^2 + \vert
B\vert^2 \right) } \int [dg] ~\otimes_i
g_{\lambda_i}(x_i)\e^{-I[g] + \frac{k}{4\pi}
\int\left\{A_- B_+^g - i B_+ \del g g^{-1} \right\}}
\ee
 The representations $\lambda_i$ act on vector spaces
$V_i$, so the wave functional is now an operator on the
tensor product of the corresponding representation
spaces,
\be
 \Xi_{\{\lambda_i\}}[\{x_i\};A,B] ~\in~ \otimes_i ~(V^*_i
\otimes V_i) \ .
\ee
 Under gauge transformations, these functionals
transform as
\be
 \Xi_{\{\lambda_i\}}[\{x_i\};A^h,B^l] = \e^{
i \alpha[B,l] -i \alpha[A,h]  }~ \otimes_i
\left(l^{-1}\right)_{\lambda_i}(x_i)
~~\Xi_{\{\lambda_i\}}[\{x_i\};A,B] ~~\otimes_i~
h_{\lambda_i}(x_i) \ .
\ee
 Moreover, the property of being annihilated by $E$ is
clearly insensitive to insertions of $g$, so these
states are indeed vacuum states for our strongly coupled
gauge theory in the presence of external particles.

With the same methods that were used to prove
(\ref{factAB}), it is easily seen that the normalization
integral for this wave functional reproduces the G/G
partition function with character insertions,
\bea
{\textstyle \frac{1}{{\rm vol}~{\cal G}}}  ~\Tr\int [dA][dB]
~~\Xi^*_{\{\lambda_i\}}[\{x_i\};A,B]
~~\Xi_{\{\lambda_i\}}[\{x_i\};A,B] ~~~~~~~~~~~~~~~~~~~~~\cr
= \int[dA][dg] ~\prod_i
\X_{\lambda_i}\left(x_i\right) ~\e^{-k~S_{{\rm
G/G}}[g,A]} \ . ~~~
\eea
 The trace is over all the representation spaces
$V^*_i \otimes V_i$.  The Verlinde formula then tells
us, which combinations of insertions give wave
functionals $\Xi_{\{\lambda_i\}}[\{x_i\};A,B]$ with
non-zero norms, and which are the true dimensions of the Hilbert
spaces spanned by the states $\Xi$.

Here we are interested in insertions in the conjugate
adjoint representation. These transform conjugate
covariantly, which counteracts the covariant
transformation of the electric field ``creation
operator'' $E^\dagger$. For concreteness, we consider
insertions of matrices
\be
 \left(g_{{\rm Ad}}^T\right)^{ab} = 2 ~\Tr\left(g T^b g^{-1} T^a\right)
\ee
 at the points $x_i$, where the trace, the generators
and the group element $g$ are in the fundamental
representation. Denoting a wave functional with $m$ such
insertions by $\Xi_m^{\{b_i,a_i\}}[\{x_i\};A,B]$, we can
act with the creation operators $E^{b\dagger}$ at the
insertion points to acquire an excited state that
is gauge invariant up to a projective phase:
\[
 \Psi_m^{\{a_i\}}[\{x_i\};A,B] =
{\textstyle \left(\frac{4\pi i}{k}\right)^m}
\prod_{i=1}^m
E^{b_i\dagger}(x_i) ~~\Xi_m^{\{b_i,a_i\}}[\{x_i\};A,B]
 ~~~~~~~~~~~~~~~~~~~~~~~~~~~~~~~~~~~
 \]

\vskip\arrayskp \be
 =
 \e^{-\frac{k}{8\pi} \int\left(\abs{A}^2 + \abs{B}^2 \right)}
 \int [dg] \prod_{i=1}^m \left(B_+
- A_+^{g^{-1}} \right)^{a_i}(x_i)
 ~\e^{-I[g] + \frac{k}{4\pi} \int\left\{A_-
B_+^g - i B_+ \del g g^{-1} \right\}} \ .
 \mabel{states}
\ee
 These states are tensor products of vectors in the adjoint
representation space $V_{{\rm Ad~}}$, or equivalently, of fundamental
representation matrices, $\Psi_m =
\Psi_m^{\{a_i\}}\otimes_{i=1}^mT^{a_i} \in \otimes_{i=1}^m V_{{\rm
Ad}}$ .
 By construction, they are eigenstates of the
Hamiltonian $H= \frac{e^2}{2} \int E^\dagger(\vec x) E(\vec
x)$. The energies of these excited states are those of Landau levels,
\be
 H ~\Psi_m[\{x_i\};A,B]
~=~ m \frac{e^2 k}{4\pi} ~\Psi_m[\{x_i\};A,B] \ ,
\ee
 and there is a continuous degeneracy labeled by the locations of the
insertions, $\{x_i\}$. These Landau levels thus consist of $m$ gauge
invariant combinations of external particles and glue, situated at the
points $x_i$.

The connection of these exited states to the ones found in
Refs. \cite{karnai} is not entirely straightforward.  There, WZNW
theory was used to describe the Jacobian arising from the change of
variables (\ref{Apm}), for the strong coupling theory with $k=0$. It
is well known \cite{pw} that this Jacobian gives rise to a WZNW theory
with level $c_v$, related to the renormalization of $k$. 
Within this framework, the authors of
\cite{karnai} constructed exited states depending of the gauge
currents. These states should be related to the ones found here, upon
eliminating the extra gauge field $B$.  In our formalism, the shift in
the energy level can be understood as arising from the short distance
singularity in $E^2$, see e.g. \cite{gawedzki}.

\bigskip

The inner product of two states of the form (\ref{states}) can be
constructed directly by tracing over the representation spaces
$V_{{\rm Ad}}$ and integrating over the gauge fields $A$ and $B$.
Summing over permutations $P$ of the positions $x_i$ of the $m$
external particles, we have
\[
 \left(\Psi_m[\{x_i\}], \Psi_n[\{y_i\}]
\right) = \delta^{mn}  \sum_{P_m} \int [dA][dB][dg][dh]
 ~\e^{-I[g] -I[h] }
~~~~~~~~~~~~~~~~~~~~~ ~~~~~~~~~~~~~~~~~~~~~~~~~~~~~
~~~~~~~~~~~~~~~~~~~~~ ~~~~~~~~~~~~~~~~~~~~~~~~~~~~~
\] \vskip \arrayskp \bea
 \times~ \exp\left\{
-\trac{k}{4\pi} \int \left[\abs{A}^2 + \abs{B}^2 - B_-A_+^g -
B_+ A_-^{h^{-1}} + iA_+\del g g^{-1} - iA_-h^{-1}\bar\del h
\right]\right\}  \cr
 ~\times~ \prod_{i=1}^m   \Tr\left[ \Bigl(B_- - A_-^g \Bigr)(x_i)~
 \Bigl(B_+ - A_+^{h^{-1}} \Bigr)(y_{P(i)}) \right]
 ~~ ~\mabel{inner}
\eea
 where the inner product of the vectors in $V_{{\rm Ad}}$ is
normalized to coincide with the fundamental representation trace.
This Gaussian path integral is easily performed. The most transparent
way is to introduce the change of variables
\be
 B_+ \to h B_+ h^{-1} + A_+^g~~,~~~
 B_- \to h B_- h^{-1} + A_+^{h^{-1}}
\ee
 with unit Jacobian.  After performing this, the
inner product can be expressed in terms of Gauged WZNW actions
$S_{G/G}[g,A]$ and gauge currents $J(g) = A^g - A$,
\[
  \left(\Psi_m[\{x_i\}], \Psi_n[\{y_i\}]
\right)
~~~~~~~~~~~~~~~~~~~~~~~~~~~~~~~~~~~~~~~~~~~~~~~
~~~~~~~~~~~~~~~~~~~~~~~~~~~~~~~~~~~~~~~~~~~~~~~
\] \vskip \arrayskp \bea
= \delta_{mn} \times \sum_{P_m}\int [dA][dB][dg][dh]
 \prod_{i=1}^m \Tr\left[ \Bigl(B_- - J_-(gh) \Bigr)(x_i)~
 \Bigl(B_+ + J_+(gh) \Bigr)(y_{P(i)}) \right] \cr
 ~\times~\exp\left\{-S_{G/G}[g,A] - S_{G/G}[h,A]
 + \trac{k}{4\pi}\int\left( J_+(g)~J_-(h^{-1}) ~-~ \abs{B}^2
\right)  \right\}
\eea
 Now we can use the gauge-generalized Polyakov-Wiegmann
formula
\be
  S_{G/G}[gh,A] =  S_{G/G}[g,A] + S_{G/G}[h,A]
 - \trac{k}{4\pi}\int J_+(g)~J_-(h^{-1})
\ee
 and the invariance of the Haar measure to transform away the
$h$ integration. The inner product reduces to the unperturbed
G/G expectation value
\be
  \left(\Psi_m[\{x_i\}], \Psi_n[\{y_i\}]
\right) =
{\textstyle {\rm vol}~{\cal G}}~~
\delta_{mn}
 \left\langle \Z_m\Bigl(\{x_i\},\{y_i\}\Bigr) \right\rangle_{G/G} ~,
\ee
  where
 \bea
 \Z_m\Bigl(\{x_i\},\{y_i\}\Bigr)
= \sum_{P_m}
 \int [dB] ~\e^{-\frac{k}{4\pi}\int\abs{B}^2}
 ~\prod_{i=1}^m\Tr\left[ \Bigl(B_- - J_- \Bigr)(x_i)~
 \Bigl(B_+ + J_+ \Bigr)(y_{P(i)}) \right] 
 \mabel{zetem}
\\
= \sum_{P_m}\sum_{l=0}^m \left(\trac{4\pi}{k}\right)^l
 \sum_{ \{x_s\}_{s=1}^l \subset  \{x_i\}_{i=1}^m}
 \!\!\!\!\!\!({\scriptstyle {\rm  dim}~G})^l
\prod_s  \delta(x_s-y_P(s)) \prod_{r\neq s}
 \Tr\left[ -J_-(x_r) J_+(y_{P(r)}) \right]
\nonumber
\eea

The expectation  values of G/G currents can be reduced to
expectation values of the  operators $R$ of
Eq. (\ref{rab}),
\be
 \left\langle
 \prod_{i=1}^m J_-^{a_i}(x_i)~J_+^{b_i}(y_i)
\right\rangle_{G/G} =
 \sum_{P} \left(\trac{8\pi}{k}\right)^m
 \left< \prod_{i=1}^m
R^{a_i b_{P(i)}}(x_i) \right>_{G/G} \delta(x_i-y_{P(i)})  \ ,
\ee
 Remembering that $\sum_a R^{aa} = {\rm dim}~G~ - \X_{{\rm
Ad}}$, we get for the inner product
\be
  \left(\Psi_m[\{x_i\}], \Psi_n[\{y_i\}] \right) =
{\textstyle {\rm vol}~{\cal G}}~~
\delta_{mn} \left(\trac{4\pi}{k}\right)^{m}
 \left\langle
\prod_{i=1}^m \X_{{\rm Ad}}(x_i)
\right\rangle_{G/G}
\sum_{P_m}  \prod_{i=1}^m \delta(x_i-y_{P(i)}) \ .
 \mabel{compinner}
\ee
 The states $\Psi_m[\{x_i\}] $ are thus orthogonal, and delta-function
normalizable with the norm $\left(\frac{4\pi}{k}\right)^{m/2}
\sqrt{{\rm vol}~{\cal G}}$. The $G/G$ expectation value of the
characters again expresses the dimensionality of the Hilbert space
spanned by the components of the wave functional $\Psi_m$, which is
the same as the dimensionality of the space spanned by the underlying
states $\Xi_m$. These are the topological invariants given by the
Verlinde formula.

\bigskip

As the orthogonal excited states $\Psi_m$ fulfill the projective gauge
transformation property (\ref{projective}), their contribution to the
finite temperature partition function is straight forward to compute
using the results above.  To connect with the path integral
formulation of Section \ref{FTT}, it will prove most transparent if
the form (\ref{zetem}) for the inner product is used.

Summing over all amounts $m$ and positions $\{x_i\}$ of external
particles, dividing with the norm, and furthermore dividing by $m!$
to kill permutations, we get for the partition function
\be
 Z_\Psi = \Tr_\Psi \left( ~\e^{- H/T} \right)
 = \sum_{m=0}^\infty ~{\textstyle\frac{1}{m!}
\left(\frac{k}{4\pi}\right)}^m \e^{-\frac{e^2 k}{4\pi T} m} ~
\left\langle\Z_m \right\rangle_{G/G} ~,
\ee
 where
\be
 \Z_m = \int [dB] ~\e^{\frac{k}{4\pi}\int\abs{B}^2}
 \left[\Tr\int (B_- - J_-) (B_+ + J_+) \right]^m ~.
 \mabel{ahjoo}
\ee
This expression needs renormalization due to the singularities arising
from the delta-function norms. Keeping this caveat in our mind, we
can commence with formally computing $Z_\Psi$.  Performing the sum
over $m$ before integrating over $B$ we get
\be
 Z_\Psi= \frac{1}{\N}
\left\langle
\exp\left\{ -\trac{k}{8\pi}
 \left(\coth\trac{e^2k}{8\pi T} - 1 \right)
 ~\Tr\int
  J_+ J_- \right\}
 \right\rangle_{G/G} ~,
\ee
 with $\N$ the normalization of Eq. (\ref{norm}).  This is exactly
the full finite temperature partition function (\ref{ftpart}).
Accordingly, the states (\ref{states}) saturate the thermal
ensemble of topologically massive 2+1-dimensional gauge
theory in the strong coupling limit, and we have found all the
physical states that contribute in this limit.

On the other hand, the functional integrations in the partition
function can be computed directly, using the normalization
(\ref{compinner}) for the physical states. Due to the delta-function
normalizability, the contribution of each $\Psi_m$ diverges as
$\delta(0)^m$. To get a finite result, we have to employ the
logarithmic scaling of the coupling constant found in Section
\ref{renor}. If the coupling is scaled as $\frac{e^2 k}{8\pi T} \sim
\ln\Lambda/\mu$, the contributions of the delta-function norms are
canceled.  In terms of the rescaled $\e^2$, the partition function
evaluates to
\be
 Z = \sum_{m=0}^\infty ~{\textstyle\frac{1}{m!}}
~\e^{-\frac{e^2 (k+c_v)}{4\pi T} m} ~V^m~
 \left\langle \prod_{i=1}^m \X_{{\rm Ad}} \right\rangle_{G/G} ~,
\ee
 where $V$ is the volume of the two-manifold $\Sigma$.  Here we have
taken into account the zero-point energy subtracted from the
Hamiltonian by the normal ordering in Eq. (\ref{normalo}), by using
the fact that the corresponding short distance singularity in $E^2$ is
known to give rise to the shift $k\to k+c_v$ in WZNW states
\cite{gawedzki}.

The partition function, within a logarithmic distance from the
topological theory, thus remains computable in terms of purely
topological quantities, the Verlinde numbers.


\section{Conclusions}

We have considered the strong coupling limit of (2+1)-dimensional
topologically massive gauge theory. The Hamiltonian is just
$H=\e^2\int E^2$, and the vacuum sector, corresponding to infinite
coupling, is the one of pure Chern-Simons theory.  
Furthermore, in the functional Scr\"odinger picture, the
vacuum can be described in terms of a G/G WZNW
theory, the partition function of which is the normalization function
of the wave functional. It is well known in fact, that G/G
WZNW partition functions reproduce dimensions of canonically quantized
Chern-Simons Hilbert spaces in terms of the so-called Verlinde
formula \cite{verlinde}.

Polyakov loops in the finite temperature theory correspond
to punctures on the surface. Polyakov loop correlators can
thus be calculated as dimensions of Chern-Simons theories on
the corresponding surfaces. These, on the other hand, are
given by correlation functions of group characters in the
G/G model. Using the Verlinde formula, we can thus calculate
all vacuum correlators of Polyakov loops in strongly coupled
topologically massive gauge theory.  These results confirm
the spontaneous center-symmetry breaking characteristic of
deconfining gauge theories without fundamental matter.

By using the heat-kernel, the finite temperature partition function
can be evaluated, this takes into account low-lying excited states
at strong (but not infinite) coupling. The finite temperature
effective theory is a perturbed G/G model, where the ``level'' of
the principal sigma-model term is perturbed away from the
topological (and conformal) fixed point.

We computed the one-loop effective potential for the 
Polyakov loop operators. 
It was found that one gets finite results for
thermodynamical quantities if $e^2/T$ scales like a logarithm of
the fundamental cut-off. This indicates the domain of reliability
of the strong coupling approximation. The effects of the suppressed
magnetic field are truly negligible as long as we stay within
logarithmic distance from the infinite coupling fixed point.

The minima of the one-loop effective potential lay on the $Z_N$ center 
symmetry, indicating that $Z_N$ is broken and that the 
theory is in a deconfined phase.

Finally, we constructed the low-lying excited states. Using
holomorphic factorization of G/G models, we found them to be
non-abelian Landau levels built upon Chern-Simons vacua on Riemann
spheres with punctures. Gauge invariant states consist of external
particle insertions corresponding to the punctures, plus glue created
by the dynamic gauge field. We showed that the heat-kernel partition
function was saturated by these states, and evaluated the partition
function. With the logarithmic scaling of the coupling constant, the
finite temperature partition function reduces to a sum of
Verlinde factors with different numbers of insertions, i.e. of zero
temperature (vacuum) correlators of the Polyakov loops corresponding
to static external particles.


\section*{Acknowledgements}

We thank F. Cooper, A Dubin, R. Jackiw, I. Kogan, 
A. Niemi, A. Polyakov, and A. Zhitnitsky for conversations on 
the subject of this paper.  
O.T. acknowledges the hospitality of M. Paranjape and LPN at
Universit\'e de Montr\'eal, where part of this work was
done. Collaboration between the University of British Columbia and
University of Perugia groups was supported partly by NATO
Collaborative grant CRG 93094 and partly by I.N.F.N. .



\end{document}